\documentclass[epj]{svjour}

\usepackage{amsmath}
\usepackage{amsfonts}
\usepackage{graphicx}

\font\c=cmsy10
\DeclareMathSymbol{\C}{\mathalpha}{AMSb}{'103}
\DeclareMathSymbol{\N}{\mathalpha}{AMSb}{'116}
\DeclareMathSymbol{\R}{\mathalpha}{AMSb}{'122}
\DeclareMathSymbol{\Z}{\mathalpha}{AMSb}{'132}

\begin{document}

\title{Steady Schr\"odinger cat state of a driven Ising chain}
\author{S. Camalet}
\institute{Laboratoire de Physique Th\'eorique de 
la Mati\`ere Condens\'ee, 
UMR 7600, Universit\'e Pierre et Marie Curie, 
Jussieu, Paris-75005, France}
\date{Received: date / Revised version: date }
\abstract{
For short-range interacting systems, 
no Schr\"odinger cat state can be stable when 
their environment is in thermal equilibrium. 
We show, by studying a chain of two-level systems 
with nearest-neighbour Ising interactions, 
that this is possible when the surroundings consists 
of two heat reservoirs at different temperatures, 
or of a heat reservoir and a monochromatic field. 
The asymptotic state of the considered system can be 
a pure superposition of mesoscopically distinct 
states, the all-spin-up and all-spin-down states, 
at low temperatures. The main feature of our model 
leading to this result is the fact that 
the Hamiltonian of the chain and the dominant part 
of its coupling to the environment obey the same 
symmetry.
\PACS{{03.65.Yz}{Decoherence; open systems; 
quantum statistical methods} 
\and {03.65.Ud}{Entanglement and quantum 
nonlocality} 
\and {05.70.Ln}{Nonequilibrium and irreversible 
thermodynamics}}
}

\maketitle

\section{Introduction}

The apparent classical behavior of macroscopic objects is 
thought to find its origin in the unavoidable interaction of 
any system with its environment \cite{Z,JZ}. For example, 
non-classical correlations between quantum systems 
\cite{B,W} are expected to be fragile against this influence. 
This fragility of quantum entanglement is confirmed by studies 
showing that an initial entanglement between two independent 
open systems disappears in a finite time \cite{DH,YE,JJ}. 
As a matter of fact, when several independent systems 
interact with an environment in thermal equilibrium, 
their state generically relaxes to their uncorrelated canonical 
thermal state. Consequently, even classical correlations are 
destroyed in this situation. This is not necessarily the case 
with a non-equilibrium surroundings. For an environment 
consisting simply of two heat reservoirs at different 
temperatures, the steady state of two non-interacting systems 
can be entangled  \cite{EPJB2}. Thus, in 
this case, an initially uncorrelated state can evolve into a 
quantum-mechanically entangled one. General results have 
been obtained concerning the steady state of 
a Markovian master equation of Lindblad form \cite{K,TV}. 
Within this approach, any pure state can be asymptotically 
reached with a purely dissipative dynamics, but not all states 
are attainable if the environment influence is required to be 
local. 

Probably the most amazing predictions of quantum theory 
arise when the superposition principle is applied to 
macroscopic objects. All states in the Hilbert space of 
any system can genuinely exist. There is no a priori restriction, 
even in the case, for instance, of the Earth's center of mass. 
Quantum coherent superpositions of macroscopically 
distinguishable states were first discussed by E. Schr\"odinger 
who considered a cat in a dead-alive state \cite{S}. Here also, 
environmental degrees of freedom are thought to play 
an essential role. Under their influence, such freak states would 
decay very quickly into statistical mixtures if they would happen 
to occur \cite{CL,WM,MNS,PHPM}. Recently, Schr\"odinger cat 
states have been realized experimentally as superpositions of 
motional wavepackets of trapped ions \cite{MMKW}, microwave 
cavity coherent fields \cite{BH,DDH}, magnetic flux states of 
superconducting quantum interference devices \cite{FPCTL}, 
internal states of trapped ions \cite{SM,LW}, photons 
polarizations \cite{ZCZYBP}, nuclear spins states in benzene 
molecules \cite{LK}, free-propagating light coherent states 
\cite{OTLG}, polarization and spatial modes of photons 
\cite{GP}. In all these experiments, the dissipative influence 
of the environment tends to destroy the created superposition 
of states and is one of the main obstacles to overcome 
to produce and observe it.

\begin{figure}
\centering
\includegraphics[width=0.45\textwidth]{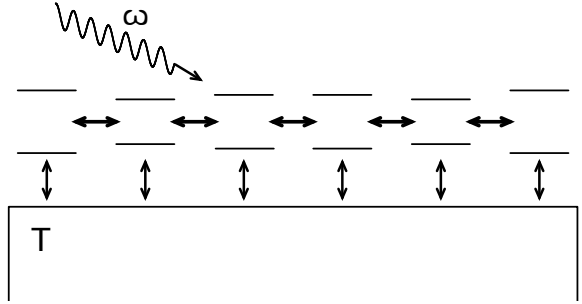}
\caption{Schematic representation of a chain of two-level systems 
with nearest-neighbour interactions and coupled to a heat reservoir 
of temperature $T$ and to a monochromatic field of frequency 
$\omega$.}
\label{fig:fig} 
\end{figure}

However, considering the positive impact on quantum 
entanglement of driving the environment out of equilibrium 
\cite{EPJB2}, one can wonder whether a Schr\"odinger cat 
state can be stable in a multiple-heat-reservoir surroundings 
or in the presence of a monochromatic field. We address 
this issue in this paper by studying a chain of two-level 
systems (TLS), with nearest-neighbour Ising interactions, 
coupled to a heat bath and a monochromatic field, 
as illustrated in Fig.\ref{fig:fig}. Couplings between 
the TLS are necessary to obtain a steady Schr\"odinger 
cat state \cite{TV}. It is proved below that, for any TLS 
system, its ground state cannot be a Schr\"odinger cat state 
when only short range interactions are present. Consequently, 
such a superposed state is necessarily unstable with 
an environment in thermal equilibrium. We will see that, 
on the contrary, if the surroundings includes a monochromatic 
field, or a second heat bath, the TLS steady state can be 
a pure superposition of mesoscopically distinct states, even 
if the interactions between the TLS are short-ranged.   

The paper is organized as follows. The model we consider is 
presented in the next section. Section~\ref{sec:TLSHs} is 
devoted to the study of the TLS Hamiltonian. In the following 
section, we derive equations that determine the asymptotic 
state of the TLS in the regime of weak coupling to both 
the heat reservoir and the monochromatic field. 
In section~\ref{sec:Scr}, we show that, when the field 
frequency is equal to a particular transition frequency of 
the TLS system, the TLS asymptotic state is, at low bath 
temperature, a steady multipartite entangled state that 
cannot be obtained at thermal equilibrium. Moreover, 
a TLS-field coupling strength regime can exist, where 
it is a pure Schr\"odinger cat state. In the last section, 
we summarize our results and mention some questions 
raised by our work. The case of a two-heat-reservoir 
environment is considered in the last Appendix.

\section{Model}

We consider a system consisting of $2N$ two-level systems, 
a heat bath and a monochromatic field, described by 
the Hamiltonian 
\begin{multline}
H = H_{TLS}  + \alpha^{-1} \epsilon_f \sigma^x_N (a^\dag+a)
+ \omega a^\dag a \label{H} \\  
+  e \sum_{n=1}^{2N}  \sigma^x_n \pi + 
\epsilon \sum_{n=1}^{2N} \left( \sigma^x_n \pi^x_n
+\sigma^y_n \pi^y_n+\sigma^z_n \pi^z_n \right)
 + H_{\cal B} , 
\end{multline}
where $\omega$ is the field frequency, $e$, $\epsilon \ll e$ 
and $\epsilon_f \ll e$ are energies characterizing the coupling 
strengths of the TLS to their environment, $\alpha$ is 
a dimensionless constant that will be defined below,
 $H_{\cal B}$ is the bath Hamiltonian, $\pi$ and $\pi^\nu_n$ 
 where $\nu \in \{ x,y,z \}$, are observables of the bath, and
\begin{equation}
H_{TLS}= - J \sum_{n=1}^{2N-1} \sigma^z_n  \sigma^z_{n+1} 
- \sum_{n=1}^{N}  h_n (\sigma^z_n-\sigma^z_{2N+1-n} ) 
\label{HTLS}
\end{equation}
is the TLS Hamiltonian. The coupling constant $J>0$ and 
the fields $h_n$ are assumed to obey $h_n>0$ and 
$\sum_n h_n<J/2$. The annihilation operator $a$ satisfies 
the bosonic commutation relation $[a,a^\dag]=1$. 
In the following, the notation ${ \tilde n }$ is used for 
the eigenvalues of the number $a^\dag a$. The Pauli 
operator $\sigma^z_n$ has eigenvalues $\pm 1$ and 
the corresponding eigenstates are denoted by 
$| \pm \rangle_n$. The observables $\sigma_n^x$ and 
$\sigma_n^y$ are then defined by 
$\sigma_n^x| \pm \rangle_n=| \mp \rangle_n$ and 
$\sigma_n^y=i\sigma_n^x\sigma_n^z$. 
The main assumption underlying our model is that 
the dominant contribution to the interaction between 
the TLS and their environment is a uniform coupling to 
the heat reservoir, of the form $\pi S_x$ where 
$S_x=\sum_{n}  \sigma^x_n$. The term proportional 
to the energy $\epsilon$ in \eqref{H}, describes 
the small deviation, which is assumed local and generic, 
of the TLS-bath interaction from this ideal form. 
The coupling to the monochromatic field can be generalized to 
$\alpha^{-1} \epsilon_f \sum_{n,\nu} \sigma^\nu_n 
(\lambda_{n\nu}a^\dag+\lambda_{n\nu}^*a)$. Our main conclusions 
remain the same provided $\lambda_{Nx}-i\lambda_{Ny}
-\lambda_{N+1x}-i\lambda_{N+1y} \ne 0$. 

The initial state of the total system is
\begin{equation}
\Omega = \sum_{k,l} r_{kl} | k \rangle \langle l |
\otimes Z^{-1} e^{ -H_\mathrm{\cal B}/T}
\otimes |\alpha \rangle \langle \alpha | ,
 \label{Omega}
\end{equation}
where $k$ and $l$ run over the eigenstates of $H_{TLS}$, 
the density matrix elements $r_{kl}$ are arbitrary, 
$T$ is the temperature of the heat bath, 
$Z=\mathrm{Tr}\exp(-H_\mathrm{\cal B}/T)$, and 
$|\alpha \rangle$ is a coherent state of the field, i.e., 
$a|\alpha \rangle=\alpha|\alpha \rangle$. 
Throughout this paper, units are used in which $\hbar=k_B=1$. 
We assume that the average boson number $\alpha^2$ is large. 
A time-dependent unitary transformation allows to change 
the Hamiltonian \eqref{H} into 
$H+2\epsilon_f \cos(\omega t) \sigma^x_N$ where $t$ is the time, 
and the initial state $|\alpha \rangle$ into the field vacuum state. 
The time evolution of the TLS with these transformed Hamiltonian 
and initial field state, is identical to that ensuing from \eqref{H} 
and \eqref{Omega} \cite{CDG}. The behavior we will find in 
the following, can thus be also obtained with a classical force 
in place of the quantum field. Another possible physical 
interpretation of the large $\alpha$ limit, with the factor 
$\alpha^{-1}$ in the Hamiltonian \eqref{H}, is as follows. 
The strength of a local coupling between the TLS $N$ and 
a cavity mode is proportional to $V^{-1/2}$ where $V$ 
is the volume of the cavity, and hence, for a given energy 
density $\omega \alpha^2 V^{-1}$, to $\alpha^{-1}$. 
Thus, large $\alpha$ means large cavity with finite 
energy density \cite{CDG}. The thermal average 
values of the bath 
operators appearing in \eqref{H} are assumed to vanish, i.e., 
$\mathrm{Tr} (\Omega \pi)=\mathrm{Tr} (\Omega \pi_n^\nu)=0$. 
This is the case, for example, for baths of spins or fermions 
in their high-temperature phases \cite{PRB1,PRL,PRB2,condmat}, 
or for heat reservoirs consisting of harmonic oscillators linearly 
coupled to the TLS \cite{LCDFGZ}. In this last example, the bath 
Hamiltonian reads $H_{\cal B}=\sum_q \omega_q a^\dag_q a_q$ 
where $[a_q,a^\dag_{q'}]=\delta_{qq'}$, and the bath observables 
$\pi_n^\nu$ and $\pi$ are linear combinations of the operators 
$a_q$ and $a^\dag_{q}$, and hence, the above mentioned 
average values clearly vanish.

\section{TLS Hamiltonian}\label{sec:TLSHs}

With the assumptions $J>0$, $h_n>0$ and $\sum_n h_n<J/2$, 
$E_0=-J (2N-1)$ is the ground energy of the TLS system and 
its degeneracy is $2$. The states
\begin{eqnarray}
&&|  \Uparrow \; \rangle = 
\mathop {\hbox{\c\char'012}} \limits_{n=1}^{2N}
| + \rangle_n ,
|  \Downarrow \; \rangle =
\mathop {\hbox{\c\char'012}} \limits_{n=1}^{2N} 
| - \rangle_n  . \label{def}
\end{eqnarray}
constitute a basis of the corresponding eigenspace. 
More generally, the configuration states 
$|  \mbox{\large $\eta$} \rangle 
= \mathop {\hbox{\c\char'012}}_{n}| \eta_n \rangle_n$ 
where $\eta_n \in \{ +,- \}$ and 
$\mbox{\large $\eta$}=(\eta_1, \ldots, \eta_{2N})$, 
are eigenstates of $H_{TLS}$. 
An important property of this Hamiltonian is that 
it commutes with the symmetry operator $\mathrm{\Pi}$ 
defined by
\begin{equation}
\mathrm{\Pi} | \mbox{\large $\eta$} \rangle 
= \mathop {\hbox{\c\char'012}} \nolimits _{n}
| \; \overline{\eta_{2N+1-n}} \; \rangle_n  ,
\label{Pi}
\end{equation}
where $\overline{\; \pm \;}=\mp$. For example, 
for $N=2$, 
$\mathrm{\Pi} {| +++- \rangle}={| +--- \rangle}$ and 
$\mathrm{\Pi} {| ++-- \rangle}={| ++-- \rangle}$, with 
a simplified notation. Clearly, $\mathrm{\Pi}^2=1$ 
and $\mathrm{\Pi}$ has only two eigenvalues, 
$1$ and $-1$. The Schr\"odinger cat states
\begin{equation}
|\mathrm{Scs}^\mp \rangle = 2^{-1/2} 
\left( |  \Uparrow \; \rangle 
\mp |  \Downarrow \; \rangle \right) \label{Scs}
\end{equation}
are ground states of $H_{TLS}$ and eigenstates of $\mathrm{\Pi}$. 
For $N=2$ and general values of the coupling constant $J$ and 
fields $h_n$, $H_{TLS}$ has ten eigenvalues $E_k$. Four are 
nondegenerate and correspond to configuration states 
which are eigenstates of $\mathrm{\Pi}$. 
The six other ones are doubly degenerate. 
For larger values of $N$, there are 
levels with higher degeneracies. For example, for $N=3$, 
${| -+---- \rangle}$, ${| -+++-- \rangle}$, ${| ++++-+ \rangle}$ 
and ${| ++---+ \rangle}$ are eigenstates of $H_{TLS}$ with 
the same eigenenergy $-J-2h_2$. 

\begin{figure}
\centering
\includegraphics[width=0.45\textwidth]{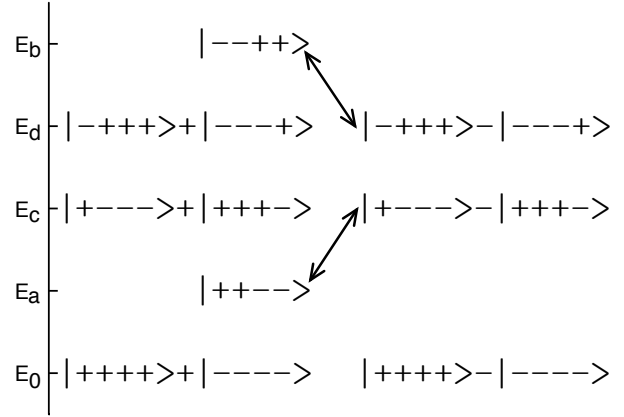}
\caption{Lowest lying eigenstates of both the TLS 
Hamiltonian $H_{TLS}$ and the symmetry operator 
$\mathrm{\Pi}$ for $4$ two-level systems. 
The eigenenergies are $E_0=-3J$, 
$E_a =-J-2h_1-2h_2$, $E_c = -J-2h_1$, $E_d=-J+2h_1$ 
and $E_b=-J+2h_1+2h_2$. The states on the left (right) 
correspond to the eigenvalue $1$ ($-1$) of $\mathrm{\Pi}$. 
The arrowed lines connect the 'left' and 'right' states 
which are coupled by the monochromatic field.}
\label{fig:spec} 
\end{figure}

Specific energy levels will play an important role in the following. 
Consider a configuration $\mbox{\large $\eta$}$ with a single interface, 
i.e., such that $\eta_n=\eta$ for $n$ smaller than a given integer $m<2N$, 
and $\eta_n=\overline{\eta}$ for $n>m$. The corresponding eigenergy 
is $E_k=E_0+2J- 2 \eta \sum_{n=1}^{m'} h_n$ where $m'=m$ if $m<N$ 
and $m'=2N-m$ otherwise. 
For general values of $J$ and $h_n$, the degeneracy of such a level is, 
for any $N$, $1$ if $m=N$ and $2$ otherwise. 
These levels are the lowest ones, see Fig.\ref{fig:spec}. 
In this figure and in the following, we use the notations 
\begin{eqnarray}
&| a \rangle = |+ \rangle_1 \ldots  | - \rangle_{N+1} \ldots , 
| b \rangle = |- \rangle_1 \ldots  | + \rangle_{N+1} \ldots ,& \label{abcd} \\
&| c^\pm \rangle = 
{\cal I}_\pm |+ \rangle_1 \ldots  | - \rangle_{N} \ldots , 
| d^\pm \rangle = 
{\cal I}_\pm |- \rangle_1 \ldots  | + \rangle_{N} \ldots ,& \nonumber
\end{eqnarray}
where ${\cal I}_\pm=2^{-1/2} (1\pm \mathrm{\Pi})$ and $\ldots$ means 
that the state of TLS $n$ is that of TLS $n-1$. These states are 
eigenstates of both $H_{TLS}$ and $\mathrm{\Pi}$. 
The corresponding eigenenergies are denoted by 
$E_a$, $E_b$, $E_c$ and $E_d$.

\section{TLS asymptotic state}\label{sec:TLSas}

For a large average boson number $\alpha^2$, there exists 
a time regime where its variation is negligible and the TLS chain 
reaches an asymptotic state $\rho_\infty$ \cite{CDG}. 
To determine $\rho_\infty$, we first write 
the reduced density matrix of the system made up of the TLS and 
the monochromatic field, at positive times $t$, as 
\begin{equation}
\rho_{TLS+F}(t) = \frac{i}{2\pi} \int_{\R+i\xi} dz  e^{-izt}  
\mathrm{Tr}_{\cal B} \left[  \left( z-{\cal L} \right)^{-1} \Omega \right]  
\label{rhot} 
\end{equation}  
where $\mathrm{Tr}_{\cal B}$ denotes the partial trace over the heat bath, 
$\xi$ is a positive real number, and the Liouvillian ${\cal L}$ is defined 
by ${\cal L} \ldots =[H,\ldots]$. The matrix elements 
${\hat r}_{k {\tilde n} l {\tilde s}}(z) = {\langle k | \langle {\tilde n} | 
\mathrm{Tr}_{\cal B} [  ( z-{\cal L} )^{-1} \Omega ]  
| l \rangle | {\tilde s} \rangle}$ of the Laplace transform of $\rho_{TLS+F}$, 
can be written as
\begin{equation}
{\hat r}_{k{\tilde n}l{\tilde s}}(z) = \sum_{k',{\tilde n}',l',{\tilde s}'} 
\Gamma_{k{\tilde n}l{\tilde s},k'{\tilde n}'l'{\tilde s}'} (z) r_{k'l'} 
\frac{\alpha^{{\tilde n}'+{\tilde s}'}}{\sqrt{{\tilde n}' !{\tilde s}' !}} 
e^{-\alpha^2}
\label{Meq}
\end{equation}  
where the functions
$\Gamma_{k{\tilde n}l{\tilde s},k'{\tilde n}'l'{\tilde s}'} (z)$ depend only 
on the heat bath part of the initial state \eqref{Omega} \cite{EPJB2}. 
The right hand side of this equation can be read as the product of 
a square matrix ${\bf \Gamma} (z)$ with a column vector. A master 
equation for $\rho_{TLS+F}$ can then be derived with the help of 
the inverse matrix ${\bf \Sigma}={\bf \Gamma}^{-1}$, see Appendix 
\ref{app:Mme}. 

We are concerned with the limit of weak coupling of the TLS to 
their environment and with $\epsilon, \epsilon_f \ll e$. It is only in 
the limit of weak coupling to its surroundings, that the density matrix 
of an open system with Hamiltonian ${\cal H}$, relaxes to the canonical 
equilibrium state $\rho_{eq} \propto \exp(-{\cal H}/T)$ when 
its environment is in equilibrium with temperature $T$. 
To obtain $\rho_\infty$ in the regime of interest, 
the expansion of ${\bf \Sigma}$ to first order in $\epsilon_f$ 
and to second order in $e$ and $\epsilon$, is required. We find 
${\bf \Sigma}={\bf \Sigma}^{0}+{\bf \Sigma}^{TLS\;{\cal B}}
+\alpha^{-1} \epsilon_f {\bf \Sigma}^{TLS\;F}+\ldots$ where
\begin{eqnarray}
\Sigma^{0}_{k{\tilde n}l{\tilde s},k'{\tilde n}'l'{\tilde s}'} =&
[z-\omega_{kl}+({\tilde s}-{\tilde n})\omega] \delta_{k'k}\delta_{l'l}
\delta_{{\tilde n}'{\tilde n}} \delta_{{\tilde s}'{\tilde s}} \label{Stls0} \\
\Sigma^{TLS\;{\cal B}}_{k{\tilde n}l{\tilde s},k'{\tilde n}'l'{\tilde s}'} =&
-i \gamma_{kl,k'l'} \big(z+({\tilde s}-{\tilde n})\omega \big) 
\delta_{{\tilde n}'{\tilde n}} \delta_{{\tilde s}'{\tilde s}} \label{Stlsb} \\
\Sigma^{TLS\;F}_{k{\tilde n}l{\tilde s},k'{\tilde n}'l'{\tilde s}'} =&
\delta_{k'k}\sigma_{l'l}\delta_{{\tilde n}'{\tilde n}}  
 \sqrt{{\tilde s}' \delta_{{\tilde s}'{\tilde s}+1}
+{\tilde s} \delta_{{\tilde s}'{\tilde s}-1} } \;  \label{Stlsf} \\
&- \delta_{l'l}\sigma_{kk'}\delta_{{\tilde s}'{\tilde s}}  
\sqrt{{\tilde n}\delta_{{\tilde n}'{\tilde n}-1}+
{\tilde n}'\delta_{{\tilde n}'{\tilde n}+1} } . \nonumber
\end{eqnarray}
In these expressions, the notations $\omega_{kl}=E_k-E_l$ and 
$\sigma_{kl}=\langle k | \sigma_N^x | l \rangle$ have 
been used. The functions $\gamma_{kl,k'l'}$, which describe 
the influence of the heat bath, can be expressed in terms of 
bath time-dependent correlation functions, see 
Appendix \ref{app:Hbi} \cite{EPJB2,QDS,CDG}. 
For a large number $\alpha^2$ and a finite time $t$, the matrix element 
${\langle k | \langle {\tilde n} | \rho_{TLS+F} | l \rangle | {\tilde s} \rangle}$ 
is non-negligible only if ${\tilde n}$ and ${\tilde s}$ are close to $\alpha^2$. 
Consequently, these integers can be replaced by $\alpha^2$ in \eqref{Stlsf}. 
Since they appear in \eqref{Stls0} and \eqref{Stlsb} only via the difference 
${\tilde s}-{\tilde n}$, it is convenient to define 
\begin{equation}
{\hat r}^{(p)}_{kl} (z) = \sum_{{\tilde n}} {\hat r}_{k{\tilde n}l{\tilde n}+p} (z) .
\label{rpkl}
\end{equation}
For $p=0$, this expression gives the matrix elements of the Laplace 
transform of the TLS chain state
\begin{equation}
\rho(t)=\mathrm{Tr}_F \rho_{TLS+F}(t) \label{TLSrho}
\end{equation}
where $\mathrm{Tr}_F$ denotes the partial trace over 
the monochromatic field. 

As we are interested in the TLS asymptotic state, we write 
the column vector ${\bf r}(z)$ of elements 
${\hat r}_{k{\tilde n}l{\tilde s}}(z)$ as
${\bf r}(z)=\sum_x {\bf r}_x/(z-x)+{\bf \tilde r}(z)$ where $x$ is real 
and ${\bf \tilde r}(z)$ has no singularity on the real axis. 
The potential pole at $z=x$, gives an undamped component 
of the TLS state, of frequency $x$, and hence contributes 
to $\rho_\infty$. Since the elements of the vector 
${\bf \Sigma}(z){\bf r}(z)$ are equal to matrix elements 
of the TLS+field initial state, it is non-singular, and thus 
the vector ${\bf r}_x$ must fulfill ${\bf \Sigma}(i0^++x){\bf r}_x=0$. 
Let us focus on the term $x=0$, which always exists, and denote 
by ${\bf u}$ the zeroth order of ${\bf r}_0$ in an expansion 
in terms of $e$, $\epsilon$ and $\epsilon_f$. Similarly to \eqref{rpkl}, 
we define $u^{(p)}_{kl}$ from the elements of ${\bf u}$. From 
${\bf \Sigma}^0 (i0^+){\bf u}=0$, it ensues that $u^{(p)}_{kl}$ is 
non-vanishing only if $\omega_{kl}=p\omega$. Then, from the next 
order terms of ${\bf \Sigma}(i0^+){\bf r}_0=0$, we obtain
\begin{multline}
\epsilon_f \sum_{j} \Big[ \sigma_{jl} \big( {u}^{(p+1)}_{kj} 
+ {u}^{(p-1)}_{kj} \big) - \sigma_{kj} 
\big( {u}^{(p+1)}_{jl} + {u}^{(p-1)}_{jl} \big) \Big] \\
-i\sum_{k',l'} \gamma_{kl,k'l'} (i0^+ +p\omega) {u}^{(p)}_{k'l'} = 0 , 
\label{MeqF}
\end{multline}
where $k$, $l$ and $p$ are such that $\omega_{kl}=p\omega$ and 
${u}^{(p')}_{k'l'}=0$ if $\omega_{k'l'} \ne p'\omega$. If $\omega$ 
is chosen arbitrarily and is not equal to any $\omega_{kl}/p$ 
then the only non-vanishing ${u}^{(p)}_{kl}$ obey $p=0$ 
and $E_k=E_l$, and are solutions of equation \eqref{MeqF} 
with no monochromatic field term. In this case, the field influence 
manifests itself only through higher order corrections. In the opposite 
case, for a resonant field frequency, the dominant contribution to 
the asymptotic behavior is affected by the coupling to 
the monochromatic field. For the more general TLS-field coupling 
$\alpha^{-1} \epsilon_f \sum_{n,\nu} 
\sigma^\nu_n (\lambda_{n\nu}a^\dag+\lambda_{n\nu}^*a)$, in 
the first line of equation \eqref{MeqF}, $\sigma_{kl}$ is replaced by 
$\sum_{n,\nu} \lambda_{n\nu} \langle k | \sigma_n^\nu | l \rangle$ 
in the first and third terms, and by 
$\sum_{n,\nu} \lambda_{n\nu}^* \langle k | \sigma_n^\nu | l \rangle$ 
in the second and fourth ones.

Since ${u}^{(0)}_{kl}=0$ if $\omega_{kl} \ne 0$, the matrix elements 
$\langle k | \rho | l \rangle$ of the TLS state \eqref{TLSrho} such that 
$E_k \ne E_l$, have no steady component. Moreover, it is clear 
from the expression \eqref{Stls0}, that, for $\omega_{kl}=0$, the only 
possible undamped component of $\langle k | \rho | l \rangle$ is of 
zero frequency, and hence 
\begin{equation}
\langle k | \rho_\infty | l \rangle =  {u}^{(0)}_{kl} \label{stme}
\end{equation}
for $E_k=E_l$. Actually, the other non-vanishing ${u}^{(p)}_{kl}$ 
give also contributions to $\rho_\infty$. This can be seen as follows. 
Let us define the vector ${\bf r}_{P\omega}$ of elements 
$({\bf r}_0)_{k{\tilde n}l{\tilde s}+P}$. It satisfies 
${\bf \Sigma} (i0^+ +P\omega){\bf r}_{P\omega}=0$ 
with the approximate ${\bf \Sigma}$ discussed above, and thus, 
corresponds to a pole of ${\bf r}(z)$ on the real axis at 
$z=P\omega$. Moreover, 
$\sum_{\tilde n} ({\bf r}_{P\omega})_{k{\tilde n}l{\tilde n}}
=\sum_{\tilde n} ({\bf r}_{0})_{k{\tilde n}l{\tilde n}+P}$ 
is equal to $u_{kl}^{(P)}$ to zeroth order, and hence 
\eqref{stme} generalizes to 
$\langle k | \rho_\infty | l \rangle =  {u}^{(p)}_{kl} \exp(-ip\omega t)$ 
for $E_k=E_l+p\omega$. In Appendix \ref{app:Mme}, we derive 
from \eqref{Stls0}-\eqref{Stlsf}, a Markovian master equation 
for $\rho(t)$ which reduces to Redfield equation when $\epsilon_f=0$, 
and show that a time-periodic 
$\rho=\sum_{p} \exp(-ip\omega t) \varrho^{(p)}$ satisfies this equation 
only if $\varrho^{(p)}_{kl}=u_{kl}^{(p)}$ to lowest order in 
$e$, $\epsilon$ and $\epsilon_f$. 
In this Appendix, we also show that, in the case of a single TLS 
coupled to a zero-temperature bath, \eqref{MeqF} gives the same 
asymptotic state as that obtained from the optical Bloch equations 
in the rotating wave approximation \cite{CDG}. 
 
\section{Schr\"odinger cat regime}\label{sec:Scr}

In this section, we consider the case of a field frequency
\begin{equation}
\omega=2h_N \label{omega} =E_c-E_a=E_b-E_d , 
\end{equation}
see Fig.\ref{fig:spec}, and of a low heat bath temperature,
\begin{equation}
T \ll  h_n < J . \label{T}
\end{equation}
We will see that, in this limit, the asymptotic state $\rho_\infty$ 
is a steady state of the form
\begin{equation}
\rho_\infty = p |\mathrm{Scs}^- \rangle \langle \mathrm{Scs}^- | 
+ (1-p) | \mathrm{Scs}^+ \rangle \langle \mathrm{Scs}^+ | 
\label{rhos}
\end{equation}
where the Schr\"odinger cat states $|\mathrm{Scs}^\mp \rangle$ 
are given by \eqref{Scs}. For vanishing coupling to the monochromatic 
field, $\epsilon_f=0$, $p=1/2$ and 
$\rho_\infty=( {| \! \Uparrow \rangle}{\langle \Uparrow \! |} 
+ {| \! \Downarrow  \rangle}{\langle \Downarrow \! |})/2$ is simply 
the zero temperature thermal state corresponding to the TLS 
Hamiltonian $H_{TLS}$. For $\epsilon_f \gg \epsilon$, 
$p=1$ and the TLS asymptotic state is a pure Schr\"odinger 
cat state. At the end of this section, it is shown that, provided 
$p \ne 1/2$, the mixed state \eqref{rhos} is not as highly 
entangled but remains nevertheless multipartite entangled. 
To obtain \eqref{rhos}, we first define 
$\rho_s=\sum_{k,l} {u}^{(0)}_{kl} | k \rangle \langle l |$, which is 
the steady component of $\rho_\infty$. From \eqref{stme} and 
Cauchy-Schwarz inequality, it follows that any matrix element 
of $\rho_\infty$ satisfies 
$|\langle k | \rho_\infty | l \rangle |^2 \le \langle k | \rho_s | k \rangle 
\langle l | \rho_s | l \rangle$, and that 
$\langle k | \rho_\infty | l \rangle=\langle k | \rho_s | l \rangle$ 
if $E_k=E_l$. In the following, 
we focus on $\rho_s$ and show that its only nonvanishing 
elements in the limit \eqref{T}, are 
$\langle \mathrm{Scs}^\mp | \rho_s | \mathrm{Scs}^\mp \rangle$, 
which leads to the result \eqref{rhos}.
  
\subsection{No thermal equilibrium Schr\"odinger cat}

Before embarking on the determination of $\rho_\infty$, let us 
first show that the thermal equilibrium state $\rho_{eq}$ of 
any TLS system cannot be of the form \eqref{rhos} with 
$p \ne 1/2$, in the absence of long range interactions. 
At temperature $T$, 
$\rho_{eq} \propto \sum_k | k \rangle \langle k | \exp(-E_k/T)$ 
where $| k \rangle$ and $E_k$ are the eigenstates 
and eigenenergies of the TLS Hamiltonian ${\cal H}$. 
If the temperature is too high, $\rho_{eq}$ is a mixture of 
a large number of states. At zero temperature, $\rho_{eq}$ 
is the equal-weight mixture of the ground states of ${\cal H}$. 
Thus, in this case, $\rho_{eq}$ can be of the form \eqref{rhos} 
but with $p=1/2$ only. The thermal state is given by 
$\rho_{eq}=p {| 0 \rangle \langle 0 |} + (1-p) {| 1 \rangle \langle 1 |}$ 
with $p \ne 1/2$, only if $| 0 \rangle$ and $| 1 \rangle$ are 
the ground state and first excited state of ${\cal H}$, and are 
non-degenerate, and only in the limit $E_2 \gg T$.  We thus 
assume that ${\cal H}$ has a nondegenerate ground state 
$|  0 \rangle$. When no long range interaction is present, ${\cal H}$ 
can be decomposed as ${\cal H}={\cal H}_1+{\cal H}'$ where no 
observable of TLS $1$ appears in ${\cal H}'$ and there exists a TLS 
$n \ne 1$ which is not affected by ${\cal H}_1$. It can be proved that 
$|  0 \rangle$ cannot be ${| \mathrm{Scs}^\mp \rangle}$ by 
evaluating the average value 
$A={\langle \mathrm{Scs}^\mp | {\cal H} | \mathrm{Scs}^\mp \rangle}$ 
as follows. Since ${\langle \Downarrow \!| {\cal H}' |\!\Uparrow \rangle} 
\propto {{_1\langle} - | + \rangle_1}=0$, and similarly for ${\cal H}_1$, 
$A=[ {\langle\Downarrow \!| {\cal H} | \!\Downarrow\rangle} 
+ {\langle \Uparrow \!| {\cal H} |\!\Uparrow\rangle}]/2$. 
Consequently, $A > {\langle 0 | {\cal H} | 0 \rangle}$, 
as this last value is the minimum possible one for the average energy, 
and $|  0 \rangle$ cannot be both ${|\!\Uparrow\rangle}$ and 
${|\!\Downarrow\rangle}$. Thus, $|  0 \rangle$ is not 
$|  \mathrm{Scs}^\mp \rangle$. This result generalizes to all 
Schmidt decomposable states, see Appendix \ref{app:NSdgs}.

\subsection{Uniform coupling to the heat bath}\label{sec:Uchb}

As we are interested in the limits $\epsilon, \epsilon_f \ll e$, 
we first consider the case $\epsilon=\epsilon_f=0$. 
For these particular values, the total Hamiltonian \eqref{H} 
commutes with the symmetry operator $\mathrm{\Pi}$ 
defined by \eqref{Pi}, and there thus exist several TLS 
asymptotic states. This can be seen as follows. Consider 
that the TLS initial state 
$\sum_{k,l} r_{kl} |k \rangle \langle l |$ is 
a statistical mixture of eigenstates of $\mathrm{\Pi}$ 
with eigenvalue $1$. Since $[H,\mathrm{\Pi}]=0$, 
the TLS reduced state at any time is also 
such a mixture. The conclusion is obviously similar for 
the eigenvalue $-1$ and hence the asymptotic state cannot 
be the same for these two kinds of initial states. Note that 
this argument is valid for any coupling strength $e$. 
We remark that, though $H_{TLS}$ and 
$S_x=\sum_n \sigma^x_n$ both commute with the symmetry 
operator $\mathrm{\Pi}$, they do not have, for general 
values of $J$ and $h_n$, any common eigenvector, 
and hence that there is no decoherence-free subspace 
in the TLS Hilbert space \cite{ZR}, 
see Appendix \ref{app:Ndfs}.

For $\epsilon_f=0$, equations \eqref{MeqF} separate into 
the independent equation sets 
$\sum_{k',l'} \gamma_{kl,k'l'} (i0^+ +p\omega) {u}^{(p)}_{k'l'} = 0$, 
where $k$ and $l$ are such that $\omega_{kl}=p\omega$, and 
${u}^{(p)}_{k'l'}=0$ if $\omega_{k'l'} \ne p\omega$. 
The equation set $p=0$ gives the steady component 
$\rho_s=\sum_{k,l} {u}^{(0)}_{kl} | k \rangle \langle l |$ of $\rho_\infty$. 
Up to now, the basis set $\{ |k \rangle \}$ has not been fully specified. 
The states $|k \rangle$ are eigenstates of $H_{TLS}$, but, 
for a degenerate energy level $E_k$, infinitely many choices are 
possible. To obtain the expression of $\rho_s$, we consider 
the diagonalized form of this density matrix in the appropriate 
unknown basis set $\{ |k \rangle \}$. The equations 
\begin{equation}
\sum_{k'} \gamma_{kl,k'k'} (i0^+) {u}^{(0)}_{k'k'} = 0 \label{Uchb}
\end{equation}
where $k$ and $l$ are such that $E_k=E_l$, then determine 
both $\{ |k \rangle \}$ and the populations $u^{(0)}_{kk}$. Let us 
introduce the coefficients 
$\Upsilon_{kl}=\gamma_{kk,ll} (i0^+) \exp(-E_l/T)$. 
They obey $\Upsilon_{lk}=\Upsilon_{kl}$ and 
$\sum_k \Upsilon_{kl}=0$, see Appendix \ref{app:Hbi}, and hence, 
for any set $\{ \varphi_{k} \}$, 
\begin{equation}
\sum_{k,l} \Upsilon_{kl} \varphi_{k} \varphi_{l} 
= -\frac{1}{2} \sum_{k,l} \Upsilon_{kl} (\varphi_{k}- \varphi_{l} )^2 . 
\label{eg}
\end{equation} 
By definition of $\Upsilon_{kl}$, this sum vanishes for 
$\varphi_{k}=u_{kk}^{(0)} \times \exp(E_k/T)$. Since 
$\Upsilon_{lk}=\Upsilon_{kl}$ and $\Upsilon_{kl} \ge 0$ 
for $k \ne l$, the vanishing of \eqref{eg} is equivalent to 
$\sum_{l} \Upsilon_{kl} \varphi_{l}=0$ \footnote{The matrix 
of elements $\Upsilon_{kl}$ is real and symmetric and hence 
diagonalizable. Using \eqref{eg} and $\Upsilon_{kl} \ge 0$ 
for $k \ne l$, it can be shown that its eigenvalues are negative, 
which leads to the equivalence.}. 
As shown in Appendix \ref{app:Hbi}, for $k \ne l$, 
$\Upsilon_{kl}$ is proportionnal to 
$S_{kl}=\langle k | S_x | l \rangle^2$ when $\epsilon=0$. 
Since the right hand side of \eqref{eg} is equal to zero only if 
$\varphi_{k}= \varphi_{l} $ when $\Upsilon_{kl} \ne 0$, 
$u^{(0)}_{kk}/u^{(0)}_{ll}=
\exp(-\omega_{kl}/T)$ when $S_{kl} \ne 0$. 
This relation ensures that the equations \eqref{Uchb} 
with $k \ne l$, are satisfied, see Appendix \ref{app:Hbi}.

Thus, $\rho_s$ is determined by the equations \eqref{Uchb} 
with $k=l$, or, equivalently, by the vanishing of the sum 
\eqref{eg}. It can always be written as
\begin{equation}
\rho_s=\sum_q \frac{p_q}{z_q} \sum_{k \in{\cal E}_q} 
e^{-E_k/T} |k \rangle \langle k | , \label{rhos2}
\end{equation}
where $z_q=\sum_{k \in{\cal E}_q} \exp(-E_k/T)$ and 
$\sum_q p_q=1$. This expression satisfies \eqref{Uchb} if 
the states $|k \rangle$ and the sets ${\cal E}_q$ are 
such that $S_{kl}=0$ for any $k \in {\cal E}_q$ 
and $l \in {\cal E}_{q'}$ where $q'\ne q$. 
Obviously, as soon as there exist 
more than one set ${\cal E}_q$, the decomposition \eqref{rhos2} 
is not unique, as a new family of sets ${\cal E}_q$ can be simply 
defined by unioning sets. Thus, we assume that a set ${\cal E}_q$ 
cannot be divided into subsets ${\cal E}_{q'}$ and ${\cal E}_{q''}$ 
such that $S_{kl}=0$ for any $k \in {\cal E}_q'$ and 
$l \in {\cal E}_{q''}$. In this case, every state $|k \rangle$ 
in \eqref{rhos2}, is eigenstate of $\mathrm{\Pi}$ and the value of 
the corresponding eigenvalue depends only on the set 
${\cal E}_q$ to which $k$ belongs. This can be shown as follows. 
Consider the subspace ${\cal S}_q$ of the TLS Hilbert space, 
spanned by $\{ |k \rangle \}$ where $k \in {\cal E}_q$. It can 
always be decomposed as 
${\cal S}_q={\cal S}_{q+} \oplus {\cal S}_{q-}$ 
where ${\cal S}_{q\pm}$ is spanned by the states 
$(1 \pm \mathrm{\Pi}) {|k \rangle}$, which are eigenstates of 
both $H_{TLS}$ and $\mathrm{\Pi}$. Since 
$[S_x,\mathrm{\Pi}]=0$, 
$\langle \psi | S_x | \psi' \rangle=0$ for any 
$| \psi \rangle \in {\cal S}_{q+}$ and 
$| \psi' \rangle \in {\cal S}_{q-}$. As ${\cal E}_q$ cannot be 
subdivided, as assumed above, ${\cal S}_q={\cal S}_{q+}$ or 
${\cal S}_q={\cal S}_{q-}$. In other words, all the states 
$|k \rangle$ given by ${\cal E}_q$, are such that 
$\mathrm{\Pi}|k \rangle=\pm|k \rangle$. 
For $N=2$, the number of levels $E_k$ is small enough 
to allow the complete determination of the minimal 
sets ${\cal E}_q$. For general values of $J$ and $h_n$, 
there are only two sets, corresponding to the two eigenvalues 
of $\mathrm{\Pi}$. For $h_2=0$, there are three sets, one 
giving only the state $({| +-++ \rangle} - {| --+- \rangle} 
- {| ++-+ \rangle} + {| -+-- \rangle})/2$. 
We remark that, since this state is eigenstate of both $H_{TLS}$ 
and $S_x$, it is decoherence-free, see Appendix \ref{app:Ndfs}. 
The situation is similar for $h_1=0$ and $h_2=h_1$. 

For our purpose, we only need to know a few states $|k \rangle$ 
and the corresponding two sets ${\cal E}_q$. The two ground states 
$| \mathrm{Scs}^- \rangle$ and $| \mathrm{Scs}^+ \rangle$
are eigenstates of $\mathrm{\Pi}$ with 
eigenvalues $-1$ and $1$, respectively. 
Thus, they are given by two different sets, ${\cal E}_{-}$ and 
${\cal E}_{+}$. Consider the doubly degenerate energy levels 
$E_e$ and $E_f$ corresponding, respectively, to the states 
$| e^\pm \rangle = {\cal I}_\pm | + \rangle_1 | - \rangle_2 \ldots$ 
and 
$| f^\pm \rangle = {\cal I}_\pm | - \rangle_1 | + \rangle_2 \ldots$, 
where ${\cal I}_\pm=2^{-1/2} (1\pm \mathrm{\Pi})$ 
\footnote{For $N=2$, $e=c$ and $f=d$, see \eqref{abcd}.}. 
Since $S_{\mathrm{Scs}^\pm e^\pm}
= S_{\mathrm{Scs}^\pm f^\pm}=1$, $e^\pm$ and $f^\pm$ 
belong to ${\cal E}_{\pm}$. By evaluating the matrix elements 
of $S_x$ between the one-interface states discussed 
at the end of section \ref{sec:TLSHs}, and noting that 
the degeneracy of a one-interface level is $1$ or $2$, 
it can be shown that ${\cal E}_+$ 
gives all the states ${\cal I}_+ | \mbox{\large $\eta$} \rangle$ 
where $| \mbox{\large $\eta$} \rangle$ is a one-interface 
configuration state which is not eigenstate of $\mathrm{\Pi}$, 
$| a \rangle$, $|b \rangle$, and $| \mathrm{Scs}^+ \rangle$. 
Similarly, ${\cal E}_-$ gives all the states 
${\cal I}_- |  \mbox{\large $\eta$} \rangle$ and 
$| \mathrm{Scs}^- \rangle$.

\subsection{General coupling to the heat bath} 

The non-uniqueness of $\rho_s$ for $\epsilon=\epsilon_f=0$ 
stems from the particular form of the coupling to the environment 
in this case. If the uniformity of this coupling is broken, 
even slightly, this indeterminacy disappears, as we show here. 
For $\epsilon \ne 0$ and $\epsilon_f=0$, the potential sets 
${\cal E}_q$, appearing in the decomposition \eqref{rhos2}, 
are determined by 
$e \langle k | S_x | l \rangle \langle A | \pi | B \rangle + 
\epsilon \sum_{n,\nu}  \langle k | \sigma^\nu_n | l \rangle 
\langle A | \pi^\nu_n | B \rangle=0$ where $|A \rangle$ 
and $|B \rangle$ are any eigenstates of the bath 
Hamiltonian $H_{\cal B}$, since $\Upsilon_{kl}$  
vanishes only if this condition is satisfied, 
see Appendix \ref{app:Hbi}. In general, it is equivalent to 
$\langle k | \sigma^\nu_n | l \rangle=0$ for 
any $n$ and $\nu$. In other words, $k$ and $l$ belong to the same 
set ${\cal E}_q$ if $\langle k | \sigma^\nu_n | l \rangle \ne 0$ for 
some $n$ and $\nu$. As a consequence, $k$ and $l$ belong to 
the same set ${\cal E}_q$ if there exist ${\tilde \jmath}_1$, \ldots, 
${\tilde \jmath}_s$, $n_1$, \ldots, $n_s$, $n$, $\nu_1$, \ldots, 
$\nu_s$ and $\nu$ such that $\langle k | 
\otimes_r (\sigma_{n_r}^{\nu_r} | {\tilde \jmath}_r \rangle
\langle {\tilde \jmath}_r |) \sigma_n^\nu | l \rangle \ne 0$, which 
means that there is a bath-induced transition 
path between the states $| k \rangle$ and $| l \rangle$. Since
\begin{multline}
\sum_{j_1, \ldots, j_s} 
\Big|\langle k | \sigma_{n_1}^{\nu_1} | j_1 \rangle  
\prod_{r=2}^s  \langle j_{r-1} | \sigma_{n_r}^{\nu_r} | j_r \rangle
 \langle j_{s} | \sigma_{n}^{\nu} | l \rangle \Big| \\
\ge \Big|\langle k | \otimes_{r=1}^s \sigma_{n_r}^{\nu_r} 
\sigma_{n}^{\nu} | l \rangle \Big| ,
\end{multline}
$k$ and $l$ belong to the same set ${\cal E}_q$ if there exist 
$n_1$, \ldots, $n_s$, $n$, $\nu_1$, \ldots, $\nu_s$ and $\nu$ 
such that $\langle k | \otimes_r \sigma_{n_r}^{\nu_r} 
\sigma_{n}^{\nu} | l \rangle \ne 0$. Consider any state $| k \rangle$ 
appearing in \eqref{rhos2}. It can be expanded on the basis of 
configuration states ${| \mbox{\large $\eta$}_r \rangle} 
= \mathop {\hbox{\c\char'012}}_{n}{| \eta_{rn} \rangle_n}$ as 
${| k \rangle} = \sum_r \lambda_{r} {| \mbox{\large $\eta$}_r \rangle}$, 
which can be rewritten as ${| k \rangle} = \sum_r \lambda_{r} 
\otimes_n {(\sigma^x_{n})^{\tau_{rn}}}{| a \rangle}$ where 
${| a \rangle} = {|+ \rangle_1 \ldots  | - \rangle_{N+1} \ldots}$, and 
$\tau_{rn}$ is equal to $1/2+\eta_{rn}/2$ for $n > N$ and to 
$1/2-\eta_{rn}/2$ otherwise. As the eigenenergy $E_a$ is 
nondegenerate, $| a \rangle$ is necessarily present in \eqref{rhos2}. 
Since at least one $\lambda_{r}$ is nonzero and 
$\langle a | \otimes_n (\sigma^x_{n})^{\tau_{rn}} | k \rangle
=\lambda_{r}$ for any $r$, $a$ and $k$ belong to 
the same set ${\cal E}_q$. Thus, there is a unique set 
${\cal E}_q$ and 
$\rho_s= \rho_{eq} \propto \sum_{k} \exp(-E_k/T) 
{|k \rangle} {\langle k |}$ 
is the thermal state of the TLS system. Consequently, in the low 
temperature limit \eqref{T}, $\rho_\infty=\rho_s=
( {| \! \Uparrow \rangle}{\langle \Uparrow \! |} 
\pm {| \! \Downarrow  \rangle}{\langle \Downarrow \! |})/2$, 
which can be written as \eqref{rhos} with $p=1/2$. We remark 
that the above proof applies even if $\pi_n^y=\pi_n^z=0$, 
and that the only required assumption on the parameters of 
$H_{TLS}$ is that they are generic enough that $E_a$ is 
non-degenerate.

\subsection{Low temperature limit}\label{sec:Ltl}

In this section, we show that, for any $\epsilon_f$ and any 
${\epsilon \ne 0}$, $\langle k | \rho_s |l \rangle$ vanishes in 
the low temperature limit \eqref{T}, if $E_k=E_l$ is larger 
than $E_0+3J$. This result, together with those of section 
\ref{sec:Uchb}, lead, for $\epsilon, \epsilon_f \ll e$, to 
the form \eqref{rhos} for $\rho_\infty$. Let us first write 
$\rho_s=\sum_k u_{kk}^{(0)} |k \rangle \langle k |$ in diagonal form, 
and observe that, if $E_l<E_k$, $\gamma_{kk,ll} (i0^+)$ vanishes 
at low temperatures, see expression \eqref{Gamma}, and that, 
for any $l$ such that $E_l>E_a$, there exists $k$ such that 
$E_k<E_l$  and $\gamma_{kk,ll} (i0^+) \ne 0$. In other words, 
the upward transition rates vanish at zero temperature, and 
there is a downward transition from any excited state 
but the first one 
${| a \rangle} = {|+ \rangle_1 \ldots  | - \rangle_{N+1} \ldots}$.
This last property can be seen as follows. Except the ground states 
${|\! \Uparrow \rangle}$ and ${|\! \Downarrow \rangle}$, 
and the one-interface states ${| \pm \rangle_1 | \mp \rangle_2 \ldots}$ 
and ${\ldots | \pm \rangle_{2N-1} | \mp \rangle_{2N}}$,
any configuration state 
${| \mbox{\large $\eta$} \rangle}$ presents a sequence 
${| \pm \rangle_n} {| \mp \rangle_{n+1}}$ where $n \notin \{ 1, 2N-1\}$. 
Thus, if ${| \mbox{\large $\eta$} \rangle}$ is neither one of the states 
mentioned above, nor 
${|\pm \rangle_1 \ldots  | \mp \rangle_{N+1} \ldots}$, it contains 
a sequence 
${| \mp \rangle_{n-1} | \pm \rangle_{n}| \mp \rangle_{n+1}}$ 
where $n \notin \{ 1, 2N-1\}$, or a sequence 
${| \pm \rangle_{n-1} | \pm \rangle_{n}
| \mp \rangle_{n+1}| \mp \rangle_{n+2}}$ 
where $n \notin \{ 1,N, 2N-1\}$. In the first case, the energy of 
$\sigma_n^x {| \mbox{\large $\eta$} \rangle}$ is lower than that of 
${| \mbox{\large $\eta$} \rangle}$ of $-2J\pm2{\tilde h}_n$ 
where ${\tilde h}_n=-h_{2N+1-n}$ for $n>N$ and $h_n$ otherwise. 
In the second case, the difference between the energies of 
${| \mbox{\large $\eta$} \rangle}$ and of 
$\sigma_n^x {| \mbox{\large $\eta$} \rangle}$ is $\pm2{\tilde h}_n$, 
and that between the energies of ${| \mbox{\large $\eta$} \rangle}$ 
and of $\sigma_{n+1}^x {| \mbox{\large $\eta$} \rangle}$ is 
$\mp2{\tilde h}_{n+1}$. Consequently, since $h_n>0$, there always 
exists a configuration state 
$\sigma_n^x {| \mbox{\large $\eta$} \rangle}$ 
with one TLS flipped, of energy lower than that of 
${| \mbox{\large $\eta$} \rangle}$. This result can be 
extended to the states ${| \pm \rangle_1 | \mp \rangle_2 \ldots}$ 
and ${\ldots | \pm \rangle_{2N-1} | \mp \rangle_{2N}}$, since flipping, 
respectively, the first and the last TLS, leads to 
one of the two ground states ${|\! \Uparrow \rangle}$ or 
${|\! \Downarrow \rangle}$, and to 
${| b \rangle} = {|- \rangle_1 \ldots  | + \rangle_{N+1} \ldots}$, 
since the energy of $\sigma_N^x{| b \rangle}$ and 
$\sigma_{N+1}^x{| b \rangle}$ is $E_d=E_b-2h_N$, 
see Fig.\ref{fig:spec}. Thus, $|a \rangle$ is the only exception.
Consider now any $l \ne a$ and expand the corresponding state 
on the appropriate configuration state basis as 
$| l \rangle = \sum_ {r=1}^s  \lambda_r 
{| \mbox{\large $\eta$}_r \rangle}$. As seen above, there exists $n$ 
such that the energy $E_k$ of 
$\sigma_n^x {| \mbox{\large $\eta$}_1 \rangle}$ 
is lower than $E_l$. There are possibly $s'$ states 
$\sigma_n^x {| \mbox{\large $\eta$}_r \rangle}$ of energy $E_k$. 
The state ${|\psi \rangle} = \sum_{r=1}^{s'} 
\lambda_{r} \sigma_n^x {| \mbox{\large $\eta$}_r \rangle}$ belongs 
to the TLS Hilbert subspace spanned by the states $|k\rangle$ of 
energy $E_k$.  Consequently, there is at least one $|k\rangle$ 
such that ${\langle k | \sigma_n^x | l \rangle} 
= {\langle k | \psi \rangle} \ne 0$, and hence, such that 
$\gamma_{kk,ll} (i0^+) \ne 0$.  

We now consider equation \eqref{MeqF} with $p=0$ and $T=0$. 
Any eigenenergy $E_k$ is of the form 
$E_k=E_0+2mJ+ \sum_n h_n \tau_n$ where $\tau_n \in \{ -2,0,2 \}$ 
and $m$ is the number of interfaces of the corresponding configuration 
states. Consequently, for general values of $J$ and $h_n$, if 
$\omega_{kl}=\pm \omega=\pm 2h_N$ then $k$ and $l$ 
correspond to the same number $m$ of interfaces. 
In particular, the monochromatic field does not couple one-interface 
states to multi-interface states. Therefore, at $T=0$,
\begin{multline}
\epsilon_f \sum_{l \in {\cal A}_{H}} \Big[ \sigma_{lk} \big( {u}^{(1)}_{kl} 
+ {u}^{(-1)}_{kl} \big) - \sigma_{kl} 
\big( {u}^{(1)}_{lk} + {u}^{(-1)}_{lk} \big) \Big] \\
-i\sum_{l \in {\cal A}_{H}} 
( \gamma_{kl} {u}^{(0)}_{ll}-\gamma_{lk} {u}^{(0)}_{kk}) 
+i\sum_{l \in {\cal A}_{L}} \gamma_{lk} {u}^{(0)}_{kk} = 0 , 
\label{zeroT}
\end{multline}
where $\gamma_{kl}$ is the zero temperature limit of 
$\gamma_{kk,ll} (i0^+)$, 
${\cal A}_{H}=\{k : E_k > E_0 + 3J \}$, 
${\cal A}_{L}=\{k : E_k < E_0 + 3J \}$, and $k \in {\cal A}_{H}$. 
The solution to this equation set can 
be interpreted as the asymptotic state of a system evolving 
under the influence of a zero temperature heat bath, 
a monochromatic field, and an additional decay mechanism 
characterized by the rates $\sum_{l \in {\cal A}_L} \gamma_{lk}$. 
Equation \eqref{zeroT} leads to 
$\sum_{k \in {\cal A}_H, l \in {\cal A}_L} 
\gamma_{lk} {u}^{(0)}_{kk} = 0$. 
Since all the coefficients $\gamma_{lk}$ in this sum are positive, 
${u}^{(0)}_{kk} = 0$ if one $\gamma_{lk}$ is different from zero. 
The results obtained above show that this is the case for 
$k \in {\cal A}_{H}$ corresponding to the lowest $E_k$ given 
by this set. Thus, 
${u}^{(0)}_{kk}={u}^{(\pm 1)}_{kl}={u}^{(\pm 1)}_{lk}=0$ for 
these $k$, and, in the equations \eqref{zeroT} determining 
the other ${u}^{(0)}_{kk}$ where $k \in {\cal A}_{H}$, 
the sets ${\cal A}_{H}$ and ${\cal A}_{L}$ can be replaced 
by the sets ${\cal A}'_{H}$ and ${\cal A}'_{L}$ obtained by, 
respectively, removing these $k$ from ${\cal A}_{H}$ and 
adding them to ${\cal A}_{L}$. This new equation set leads to 
${u}^{(0)}_{kk}={u}^{(\pm 1)}_{kl}={u}^{(\pm 1)}_{lk}=0$ for 
$k \in {\cal A}'_{H}$ corresponding to the lowest $E_k$ given 
by this set. By repeating this procedure, it can be shown that 
${u}^{(0)}_{kk}=0$ for all $k \in {\cal A}_{H}$, i.e., such that 
$E_k>E_0+3J$. 

For $\epsilon,\epsilon_f \ll e$, we know that 
$\rho_s$ is given by \eqref{rhos2}. The results of this section 
show that the probabilities $p_q$ where $q \ne \pm$, vanish 
at low temperatures, since the corresponding sets ${\cal E}_q$ 
give states with more than one interface. Thus, 
the low temperature TLS asymptotic state is 
of the form \eqref{rhos}. By slightly modifying the above 
derivation, it can be shown that $p_q \ll \exp(-3J/T)$ 
where $q\ne \pm$. At low temperatures, 
$\gamma_{kk,ll} (i0^+) \simeq \gamma_{lk}\exp(-\omega_{kl}/T)$
if $E_l<E_k$, see expression \eqref{Gamma}. 
Hence, the terms neglected in equation \eqref{zeroT}, 
are approximatively given by $-i\sum_{l \in {\cal A}_L} 
\gamma_{lk} \times \exp(-\omega_{kl}/T) {u}^{(0)}_{ll}$. 
Since ${u}^{(0)}_{ll} \simeq p_\pm \exp[-(E_l-E_0)/T]$ for 
$l \in {\cal E}_{\pm}$, and ${\cal A}_L \subset {\cal E}_+ \cup {\cal E}_-$, 
the above sum is far smaller than $\exp(-3J/T)$, and hence 
$\exp(3J/T){u}^{(0)}_{kk}$ where $k \in {\cal A}_H$, vanishes in 
the zero temperature limit. This result will be useful in the following.

\subsection{Steady Schr\"odinger cat state}\label{sec:SScs}

We assume here that $\epsilon \ll \epsilon_f \ll e$. To obtain $\rho_s$ 
in this case, it is convenient to first rewrite equations \eqref{MeqF} 
in matrix form as ${\bf G}_0^{(p)}{\bf u}^{(p)}
+\epsilon_f {\bf G}_1^{(p)}{\bf u}^{(p+1)}
+\epsilon_f {\bf G}_{-1}^{(p)}{\bf u}^{(p-1)} {=0}$ where ${\bf u}^{(p)}$ 
is the column vector whose elements are the non-vanishing 
$u^{(p)}_{kl}$, i.e., such that $\omega_{kl}=p\omega$. 
In the following, we use the inverse matrices $({\bf G}_0^{(p)})^{-1}$ 
where $p \ne 0$. Their existence can be shown as follows. 
If $0$ were an eigenvalue of ${\bf G}_0^{(p)}$, possible asymptotic 
states for $\epsilon_f=0$, would be 
$\rho_\infty=\rho_{eq}+x\rho_{osc}$ 
where $\rho_{osc}$ is a matrix with only nondiagonal elements, 
oscillating at frequency $p\omega$, constructed from 
an eigenvector of ${\bf G}_0^{(p)}$ with eigenvalue $0$, and $x$ 
is any real number. But, since $\rho_\infty$ is a density matrix, 
its off-diagonal elements obey 
$x^2|\langle k | \rho_{osc} | l \rangle|^2 < 1$,
which is not possible for any $x$ if $\rho_{osc} \ne 0$. 
Thus, $0$ is not an eigenvalue of ${\bf G}_0^{(p)}$, which is 
hence invertible.  Physically, this means that the possible 
oscillating components of $\rho_\infty$ with frequencies 
$p\omega$, are induced by the monochromatic field and 
disappear when the coupling to it vanishes. 
As a consequence, the contribution to $\rho_\infty$ 
which oscillates at frequency $\omega$ has an amplitude of 
the order of $\epsilon_f$, and, more generally, ${\bf u}^{(p)}$ 
is of the order of $\epsilon_f^{|p|}$. Thus, ${\bf u}^{(0)}$ 
satisfies 
\begin{multline}
\left[ {\bf G}_0^{(0)}-\epsilon_f^2 {\bf G}_1^{(0)}
\left( {\bf G}_0^{(1)} \right)^{-1} 
 {\bf G}_{-1}^{(1)} \right. \\ 
 \left. -\epsilon_f^2 {\bf G}_{-1}^{(0)}\left( {\bf G}_0^{(-1)} \right)^{-1} 
 {\bf G}_{1}^{(-1)} \right] {\bf u}^{(0)}= 0 , \label{G}
\end{multline}
where terms of order $\epsilon_f^3$ and higher have been 
neglected. 

We denote by ${\bf \tilde G}_0^{(p)}$ the matrix ${\bf G}_0^{(p)}$ 
with $\epsilon$ set to zero. As seen in section \ref{sec:Uchb}, 
${\bf \tilde G}_0^{(0)}$ has several eigenvectors ${\bf \Psi}_q$ 
with eigenvalue $0$, whose 
elements are $z_q^{-1} \exp(-E_k/T) \delta_{kl}$ if $k \in {\cal E}_q$ 
and $0$ otherwise, with the basis set $\{|k\rangle \}$ 
considered in section \ref{sec:Uchb}. There also exist column 
vectors ${\bf \Phi}_q$ which obey 
${\bf \Phi}_{q}^T{\bf \tilde G}^{(0)}_0=0$ and 
${\bf \Phi}_{q}^T {\bf \Psi}_{q'}=\delta_{qq'}$, 
and whose elements are $\delta_{kl}$ if $k \in {\cal E}_q$ and 
$0$ otherwise, see Appendix \ref{app:Hbi}. 
To zeroth order in $\epsilon_f$ and $\epsilon$, 
${\bf u}^{(0)}=\sum_q p_q {\bf \Psi}_q$. Replacing ${\bf u}^{(0)}$ 
by this expression in \eqref{G} and neglecting terms proportional 
to powers of $\epsilon$, leads to 
$\sum_{q'} r_{q q'} p_{q'} = 0$ with
\begin{equation}
r_{qq'}={\bf \Phi}_{q}^T \left[ {\bf G}_1^{(0)}{\bf F}^{(1)} 
{\bf  G}_{-1}^{(1)} + {\bf G}_{-1}^{(0)} {\bf F}^{(-1)} 
{\bf G}_{1}^{(-1)} \right]  {\bf \Psi}_{q'} 
\label{rqq'0}
\end{equation}
where ${\bf F}^{(\pm 1)}=({\bf \tilde G}_0^{(\pm 1)})^{-1}$. 
Since the monochromatic field couples states given by 
different sets ${\cal E}_q$, see Fig.\ref{fig:spec}, 
the probabilities $p_q$ appearing in \eqref{rhos2} are no 
longer independent from each other.

With the explicit expressions of the matrices 
${\bf \tilde G}_{\pm 1}^{(0)}$ and ${\bf G}_{\mp 1}^{(\pm 1)}$, 
and of the vectors ${\bf \Phi}_{q}$ and ${\bf \Psi}_{q}$, we find
\begin{equation}
r_{qq'}  = \frac{2}{z_{q'}}\mathrm{Re} \sum_{k,l,k',l' \atop p=\pm 1} 
e^{-E_{k'}/T}\big[\delta_{qq'}\delta_{q}^{k'}
-\delta_q^l \delta_{q'}^{k'} \big] 
\sigma^{(p)}_{kl} \sigma^{(p)}_{k'l'}  {\bf F}^{(p)}_{kl,k'l'} \label{rqq'}
\end{equation}
where $\sigma^{(\pm 1)}_{kl}=\langle k | \sigma_N^x | l \rangle$ 
when $\omega_{kl}=\pm \omega$ and $0$ otherwise, and 
$\delta_q^k=1$ when $k \in {\cal E}_q$ and $0$ otherwise. 
For the more general TLS-field coupling 
$\alpha^{-1} \epsilon_f \sum_{n,\nu} \sigma^\nu_n 
(\lambda_{n\nu}a^\dag+\lambda_{n\nu}^*a)$, 
$\sigma_{kl}^{(1)}=
\sum_{n,\nu} \lambda_{n\nu} \langle k | \sigma_n^\nu | l \rangle$ 
and $\sigma_{kl}^{(-1)}=(\sigma_{lk}^{(1)})^*$, which gives 
$\sigma_{c^-a}^{(1)}=2^{-1/2} (\lambda_{Nx}-i\lambda_{Ny}
-\lambda_{N+1x}-i\lambda_{N+1y})$. 
To obtain the expression \eqref{rqq'}, we have used the fact that 
${\bf F}^{(p)}_{kl,k'l'}$ vanishes if $k$ and $k'$ do not belong 
to the same set ${\cal E}_q$, or if $l$ and $l'$ are elements 
of different sets, and 
${\bf F}_{kl,k'l'}^{(-p)}=({\bf F}_{lk,l'k'}^{(p)})^*$, 
see Appendix \ref{app:Hbi}. The coefficients \eqref{rqq'} 
satisfy $\sum_q r_{qq'}=0$. We have seen in the previous 
section that $p_q$ where $q \ne \pm$, vanishes at low 
temperatures, faster than $\exp(-3J/T)$. 
The populations $p_-$ and $p_+$ obey 
$r_{--}p_- + r_{-+} p_+ + \sum_{q \ne \pm} r_{-q} p_q=0$. 
For low temperatures, 
$r_{\pm +} \simeq \pm \exp(-E_a/T)r$ 
and $r_{\mp -} \simeq \pm \exp(-E_c/T)r$ where
\begin{equation}
r = e^{E_0/T} \mathrm{Re} 
\Big[\big(\gamma_{c^-a,c^-a}(i0^++\omega)\big)^{-1}\Big] , 
\label{r}
\end{equation}
since ${\bf F}^{(1)}_{kl,c^- a} \propto \delta_{kc^-} \delta_{la}$ 
at zero temperature, see Appendix \ref{app:Hbi}. 
Consequently, $p_+/p_- \simeq -r_{--}/r_{-+} \simeq 
\exp(-\omega/T)$ vanishes in the low temperature limit \eqref{T}, 
and thus the asymptotic 
state $\rho_\infty=|\mathrm{Scs}^- \rangle \langle \mathrm{Scs}^- |$ 
is a pure Schr\"odinger cat state. 
We remark that, at strictly zero temperature, there 
is no upward transition depopulating the ground state 
$|\mathrm{Scs}^+ \rangle$ and hence the TLS system 
does not necessarily relax into the pure state 
$|\mathrm{Scs}^- \rangle$. A steady pure Schr\"odinger 
cat state can also be obtained by coupling the TLS to a 
second heat bath instead of a monochromatic field, 
see Appendix \ref{app:Thre}.

\subsection{Steady multipartite entangled state}

As shown in the previous section, the TLS asymptotic 
state $\rho_\infty$ is a pure Schr\"odinger cat state when 
$\epsilon/\epsilon_f \ll 1$. For finite values of this ratio, 
$\rho_\infty$ is given by \eqref{rhos}, and is not such a 
pure superposition of mesoscopically distinct states. However, 
it remains multipartite entangled as long as $p \ne 1/2$. 
More precisely, there is no partition of the TLS system, with respect 
to which, it is separable. It is enough to prove it for an arbitrary 
bipartite splitting. Let us then consider such a partition, i.e., 
any two subsets of $\{ 1, \ldots , 2N\}$, and name $TLS_1$ and 
$TLS_2$ the corresponding TLS systems. The asymptotic 
state of $TLS_1$ is $\rho_1=\mathrm{Tr}_{TLS_2} \rho_\infty$, 
and that of $TLS_2$ is obtained by tracing over $TLS_1$. 
The expression \eqref{rhos} can be rewritten as 
\begin{equation}
\rho_\infty = 
\frac{1}{2}\big( |\!\! \Uparrow \rangle \langle  \Uparrow \!\!| + 
| \!\! \Downarrow \rangle \langle \Downarrow \!\!| \big)
+\Big(\frac{1}{2}-p \Big) 
\big( |\!\! \Uparrow \rangle \langle  \Downarrow \!\!| + 
| \!\! \Downarrow \rangle \langle \Uparrow \!\!| \big) , \label{rhos3}
\end{equation} 
which shows that $\rho_\tau$ is an equal-weight mixture of 
the all-spin-up and all-spin-down states of $TLS_\tau$. 
Consequently, the von Neumann entropy of $\rho_\tau$ is 
$S(\rho_\tau)=-\mathrm{Tr} (\rho_\tau \ln \rho_\tau)=\ln 2$. 
On the other hand, the entropy of the complete TLS system 
state \eqref{rhos} is $S(\rho_\infty)=-p\ln p -(1-p)\ln(1-p)$. 
For $p \ne 1/2$, $S(\rho_\tau)>S(\rho_\infty)$, and hence, 
the systems $TLS_1$ and $TLS_2$ are entangled 
\cite{HHHH}.

\section{Conclusion}

In this paper, we have seen that, for any TLS system with 
short-range interactions, no Schr\"odinger cat state can be 
stable when the system environment is in thermal equilibrium.
To examine whether this is possible when the environment is 
out of equilibrium, we have studied a chain of two-level systems 
coupled to a heat reservoir and to a monochromatic field. 
For any even number of TLS, we found a regime of Hamiltonian 
parameters where the asymptotic state of the TLS chain is 
a pure Schr\"odinger cat state at low temperatures. Though 
obtained for a specific model, this result, together with that 
of Ref.\cite{EPJB2}, suggests that, more generally, 
driving the environment out of equilibrium can 
enhance considerably non-classical features of an open system. 
It would be interesting, for other models, to study how diverse 
non-classicality criteria, based on Wigner function for example 
\cite{PHPM}, or entanglement measures, evaluated for 
the system steady state, change with the distance from 
equilibrium of the environment.   
 
The existence of a Schr\"odinger cat regime may not be 
specific to the TLS chain studied in this paper. It ensues from 
some main features other systems can present, which 
are the following. The ground level of the system is degenerate. 
This is essential since, as we have seen, a non-degenerate 
ground state cannot be a superposition of macroscopically 
distinct states. The system Hamiltonian and the dominant 
component of the interaction with the environment commute with 
the same symmetry operator. This leads to the possibility of 
unequal steady-state populations for equal-energy eigenstates 
of the system Hamiltonian. The relation between the populations 
of the ground states is determined by the non-symmetric part of 
the system-environment coupling. The presence of 
non-degenerate energy levels plays an important role in 
the existence of a regime where one of these two populations 
vanishes. It would be of interest to study other systems showing 
the same features. 
 
 \begin{appendix}

\section{Markovian master equation}\label{app:Mme}

\begin{sloppypar}
By inverting the matrix ${\bf \Gamma}(z)$ in \eqref{Meq}, 
one obtains the master equation 
\begin{equation}
\sum_{k',{\tilde n}',l',{\tilde s}'} 
\Sigma_{k{\tilde n}l{\tilde s},k'{\tilde n}'l'{\tilde s}'} (z) 
{\hat r}_{k'{\tilde n}'l'{\tilde s}'}(z) =  r_{kl} 
\frac{\alpha^{{\tilde n}+{\tilde s}}}{\sqrt{{\tilde n} !{\tilde s} !}} 
e^{-\alpha^2} 
\end{equation}  
where ${\bf \Sigma}={\bf \Gamma}^{-1}$. Expanding this matrix 
as ${\bf \Sigma}={\bf \Sigma}^{0}+{\bf \Sigma}^{TLS\;{\cal B}}
+\alpha^{-1} \epsilon_f {\bf \Sigma}^{TLS\;F}+\ldots$ where 
${\bf \Sigma}^{0}$, ${\bf \Sigma}^{TLS\;{\cal B}}$ and 
${\bf \Sigma}^{TLS\;F}$ 
are given by the expressions \eqref{Stls0}-\eqref{Stlsf} and taking 
into account that the number $\alpha^2$ is large, leads to 
\begin{multline}
\partial_t r^{(p)}_{kl} + i(\omega_{kl}-p\omega){r}^{(p)}_{kl} \\ 
-i \epsilon_f \sum_{j} \Big[ \sigma_{jl} \big( {r}^{(p+1)}_{kj} 
+ {r}^{(p-1)}_{kj} \big) - \sigma_{kj} 
\big( {r}^{(p+1)}_{jl} + {r}^{(p-1)}_{jl} \big) \Big] \\
-\sum_{k',l'} \int_0^t dt' K_{kl,k'l'} (t') e^{ip\omega t'} 
{r}^{(p)}_{k'l'} (t-t') = 0 
\label{Meqt}
\end{multline}
where $r^{(p)}_{kl}=\sum_{\tilde n} 
\langle k | \langle {\tilde n} | 
\rho_{TLS+F} | l \rangle | {\tilde n}+p \rangle$, 
which is related to \eqref{rpkl} by Laplace transform.  
The time functions $K_{kl,k'l'}$ are given by  
\begin{multline}
K_{kl,k'l'} (t)=e^{it\omega_{l'k}} C^{l'l}_{kk'} (t) 
+ e^{it\omega_{lk'}} C^{l'l}_{kk'} (-t) \\
- \sum_j \big[ \delta_{ll'} e^{it\omega_{l j}} C^{kj}_{jk'} (t) 
+ \delta_{kk'} e^{it\omega_{jk}} C^{l'j}_{jl} (-t)]  \label{K}
\end{multline}
where 
$C^{kl}_{k'l'} (t) =  \mathrm{Tr}[\Omega \pi_{kl} (t) \pi_{k'l'}]$ 
are bath correlation functions \cite{CDG,EPJB2}, and, 
with the Hamiltonian \eqref{H}, 
\begin{equation}
\pi_{kl}= e \sum_n \langle k | \sigma^x_n | l \rangle \pi + 
\epsilon \sum_{n,\nu}  
\langle k | \sigma^\nu_n | l \rangle \pi^\nu_n  ,
\end{equation}
and $\pi_{kl} (t)= \exp(-itH_{\cal B})\pi_{kl}\exp(itH_{\cal B})$.  
It is clear from the above definition that the components 
$r^{(0)}_{kl}=\langle k | \rho | l \rangle$ are the matrix 
elements of the TLS reduced density matrix 
$\rho=\mathrm{Tr}_F \rho_{TLS+F}$. 
\end{sloppypar}

For a large boson number $\alpha^2$, the time evolution 
of the monochromatic field is essentially not affected by 
its interaction with the TLS and 
$r^{(p)}_{kl} \simeq \exp(ip\omega t) r^{(0)}_{kl}$. 
Moreover, since the coupling of the TLS to the heat 
reservoir is weak, the Markovian approximation 
$r^{(0)}_{kl} (t-t') \simeq \exp(i\omega_{kl} t') r^{(0)}_{kl}(t)$ 
can be used in the right hand side of \eqref{Meqt}. These two 
approximations give the Markovian master equation 
\begin{multline}
\partial_t r_{kl} + i\omega_{kl}{r}_{kl} 
-\sum_{k',l'} \gamma_{kl,k'l'} (i0^+ +\omega_{k'l'}) {r}_{k'l'} \\
-2 i \epsilon_f\cos(\omega t) \sum_{j} \big(  \sigma_{jl} {r}_{kj}
 - \sigma_{kj} {r}_{jl} \big)  = 0 \label{Meqt2}
\end{multline}
where $r_{kl}=r^{(0)}_{kl}=\langle k | \rho | l \rangle$ and 
the functions $\gamma_{kl,k'l'}$ are given by \eqref{petitgamma}, 
which reduces to Redfield equation \cite{QDS,CDG} for 
$\epsilon_f=0$. Assuming time periodic 
$r_{kl}=\sum_p \exp(-ip\omega t) \varrho^{(p)}_{kl}$ leads to
\begin{multline}
i(\omega_{kl}-p\omega)\varrho^{(p)}_{kl} 
-\sum_{k',l'} \gamma_{kl,k'l'} (i0^+ +\omega_{k'l'}) 
\varrho^{(p)}_{k'l'} \\
-i\epsilon_f \sum_{j} \Big[ \sigma_{jl} 
\big( \varrho^{(p+1)}_{kj} 
+ \varrho^{(p-1)}_{kj} \big) - \sigma_{kj} 
\big( \varrho^{(p+1)}_{jl} + \varrho^{(p-1)}_{jl} \big) \Big] = 0 , 
\label{period}
\end{multline}
for any $k$, $l$ and $p$. We solve this equation set 
perturbatively in both the coupling to the heat bath and to 
the monochromatic field, and name $u^{(p)}_{kl}$ the zeroth 
order of $\varrho^{(p)}_{kl}$. The first term of \eqref{period} 
imposes that $u^{(p)}_{kl}=0$ if $\omega_{kl} \ne p\omega$. 
Then the lowest order of \eqref{period} for $k$, $l$ and $p$ 
such that $\omega_{kl}=p\omega$, 
gives equation \eqref{MeqF}.

For a single TLS coupled to a zero-temperature heat bath 
and to a field of frequency $\omega=2h$, described by 
the Hamiltonian $H=-h\sigma^z
+ \alpha^{-1} \epsilon_f \sigma^x (a^\dag+a)
+ \omega a^\dag a+\pi \sigma^x+H_{\cal B}$, 
\eqref{MeqF} reads
\begin{eqnarray}
\gamma u^{(0)}_{--} +i \epsilon_f 
\big( u^{(-1)}_{+-} - u^{(1)}_{-+} \big) &=& 0 \nonumber \\
- \gamma u^{(0)}_{--} +i\epsilon_f 
\big( u^{(1)}_{-+} - u^{(-1)}_{+-} \big) &=& 0 \\
- (\gamma/2+i\theta) u^{(-1)}_{+-} +i\epsilon_f 
\big( u^{(0)}_{++} - u^{(0)}_{--} \big) &=& 0 
\nonumber
\end{eqnarray}
where $\gamma=\sum_{A,B} \delta(E_A) \delta(E_B-2h) 
|\langle A | \pi  | B \rangle|^2$ 
and $\theta=
\sum_{A,B} \delta(E_A) |\langle A | \pi  | B \rangle|^2 
2\omega/(E_B^2-\omega^2)$, which give the asymptotic 
solution of the well-known optical Bloch equations in 
the rotating wave approximation. We recall that 
these equations are obtained by neglecting the non-secular 
terms of the Redfield part of \eqref{Meqt2} and the non-resonant 
coupling terms to the monochromatic field, which is valid in 
the limit of weak coupling to the heat bath and to 
the monochromatic field \cite{CDG}.

\section{Heat bath influence}\label{app:Hbi} 

The functions $\gamma_{kl,k'l'}$ which appear in \eqref{Stlsb}, 
are related to the time functions $K_{kl,k'l'}$ given by \eqref{K}, by
\begin{equation}
\gamma_{kl,k'l'} (z) = \int_0^\infty dt e^{izt} K_{kl,k'l'} (t) .
\label{petitgamma}
\end{equation} 
If the correlation functions $C^{kl}_{k'l'}$ vanish fast enough in 
the long time limit, the functions \eqref{petitgamma} are 
non-singular on the real axis where they can be written as
\begin{multline}
\gamma_{kl,k'l'} (i0^+ +\omega) = \pi \sum_{A,B} P_A \bigg\{ 
\langle A | \pi_{l'l}  | B \rangle \langle B | \pi_{kk'}  | A \rangle \\
\times \Big[ {\tilde \delta}(E_A-E_B+\omega+\omega_{l'k}) 
+ {\tilde \delta}(-E_A+E_B+\omega+\omega_{lk'}) \Big] \\
-\delta_{ll'} \sum_j \langle A | \pi_{kj}  | B \rangle 
\langle B | \pi_{jk'}  | A \rangle 
 {\tilde \delta}(E_A-E_B+\omega+\omega_{lj}) \\
-\delta_{kk'} \sum_j \langle A | \pi_{l'j}  | B \rangle 
\langle B | \pi_{jl}  | A \rangle 
 {\tilde \delta}(-E_A+E_B+\omega+\omega_{jk})  
 \bigg\} , \label{gammareel}
\end{multline} 
where ${\tilde \delta} (\omega)=\delta(\omega)
+(i/\pi)\omega^{-1}$, since $[\Omega,H_{\cal B}]=0$. 
In this expression, $E_A$ and $| A \rangle$ denote 
the eigenvalues and eigenstates of the bath Hamiltonian 
$H_\mathrm{\cal B}$, 
$P_A=\mathrm{Tr}[\Omega | A \rangle \langle A |]$ is 
the initial population of state $| A \rangle$. For the initial state 
\eqref{Omega}, $P_A=Z^{-1} \exp(-E_A/T)$. 

For $\omega=0$, $l'=k'$ and $E_l=E_k$, expression 
\eqref{gammareel} simplifies to
\begin{multline}
\gamma_{kl,k'k'} (i0^+) = \sum_{A,B} P_A \bigg\{ 
- \sum_j \langle A | \pi_{kj}  | B \rangle 
\langle B | \pi_{jl}  | A \rangle \\
\times \Big[ \pi (\delta_{lk'}+\delta_{kk'}) 
\delta(E_A-E_B+\omega_{kj}) 
+i\frac{\delta_{lk'}-\delta_{kk'}}{E_A-E_B+\omega_{kj}} \Big] \\
+ 2 \pi \langle A | \pi_{k'l}  | B \rangle 
\langle B | \pi_{kk'}  | A \rangle 
\delta(E_A-E_B+\omega_{k'k})  \bigg\} , \label{gamma0}
\end{multline} 
which leads to
\begin{eqnarray}
\Upsilon_{kl} &=& \gamma_{kk,ll} (i0^+) e^{-E_l/T} 
\label{Gamma} \\ 
&=& 2\pi \sum_{A,B} P_A e^{-E_l/T} \bigg\{ 
|\langle A | \pi_{lk}  | B \rangle |^2 
\delta(E_A-E_B+\omega_{lk}) \nonumber \\
&~&- \delta_{kl} \sum_j |\langle A | \pi_{kj}  | B \rangle|^2 
\delta(E_A-E_B+\omega_{kj}) \bigg\} . \nonumber
\end{eqnarray} 
For $P_A \propto \exp(-E_A/T)$, since $\omega_{lk}=E_l-E_k$, 
$\Upsilon_{lk}=\Upsilon_{kl}$. For $\epsilon=0$, 
$\pi_{kl}= e \pi \langle k | S_x | l \rangle$ where 
$S_x=\sum_n \sigma^x_n$, and hence, for $k \ne l$, 
$\Upsilon_{kl} \propto S_{kl}$ where 
$S_{kl}=\langle k | S_x | l \rangle^2$. 
Using \eqref{gamma0}, one finds, for $E_l=E_k$,
\begin{multline}
\sum_{k'} \gamma_{kl,k'k'} (i0^+) p_{k'} =  - \pi \sum_{A,B,k'} P_A 
\langle A | \pi_{kk'} | B \rangle \langle B | \pi_{k'l} | A \rangle \\
\times \left[ \left(p_k+p_l-2\frac{P_B}{P_A}p_{k'} \right) 
\delta(\omega_{AB}+\omega_{kk'}) 
+\frac{i}{\pi} \frac{p_l-p_k}{\omega_{AB}+\omega_{kk'}} \right]  
\label{sum}
\end{multline}
where $\omega_{AB}=E_A-E_B$. For $\epsilon=0$, the above 
summand vanishes if $S_{kk'}=0$, $S_{k'l}=0$, or 
$p_{k'}=\exp(\omega_{kk'}/T) p_k=\exp(\omega_{kk'}/T) p_l$, 
which ensures that \eqref{rhos2} is the solution to \eqref{Uchb}. 
Similarly, one finds, for $E_{l'}=E_{k'}$, 
\begin{multline}
\sum_{k} p_k \gamma_{kk,k'l'} (i0^+) =  \pi \sum_{A,B,k} P_A 
\langle A | \pi_{l'k} | B \rangle \langle B | \pi_{kk'} | A \rangle \\
\times \left[ \left(2p_k-p_{k'}-p_{l'} \right) 
\delta(\omega_{AB}+\omega_{k'k}) 
+\frac{i}{\pi} \frac{p_{k'}-p_{l'}}{\omega_{AB}+\omega_{k'k}} 
\right] . 
\end{multline}
For $\epsilon=0$, this sum vanishes if $p_k=p_l$ when 
$S_{kl} \ne 0$, which gives the left eigenvectors ${\bf \Phi}_q$ 
introduced in section \ref{sec:SScs}. 

The elements of the matrix ${\bf \tilde G}_0^{(p)}$, defined 
in section \ref{sec:SScs}, are given by
\begin{multline}
\left( {\bf \tilde G}_0^{(p)} \right)_{kl,k'l'} 
=\gamma_{kl,k'l'} (i0^+ +\omega_{kl})\\
= \pi e^2 \sum_{A,B} P_A |\langle A | \pi  | B \rangle|^2
 \bigg\{ 2 s_{l'l} s_{kk'}  \delta(\omega_{ABl'l}) \\
- \sum_j  \Big[  \delta_{ll'} s_{kj} s_{jk'} 
{\tilde \delta}(\omega_{ABkj}) 
+ \delta_{kk'} s_{l'j} s_{jl} {\tilde \delta}(-\omega_{ABlj}) \Big]  
\bigg\} , \label{G0p}
\end{multline} 
where $s_{kl}=\langle k | S_x | l \rangle$ and 
$\omega_{ABkl}=\omega_{AB}+\omega_{kl}=E_A-E_B+E_k-E_l$. 
The equalities $\omega_{kl}=\omega_{k'l'}=p\omega$ have been 
used to simplify this expression. The integer $p$ does not appear 
explicitly in \eqref{G0p}, but this expression is meaningful, 
for a given $p$, only for pairs $(k,l)$ and $(k',l')$ such that 
$\omega_{kl}=\omega_{k'l'}=p\omega$. Since $s_{kl}=0$ when 
$k$ and $l$ do not belong to the same set ${\cal E}_q$, 
the element \eqref{G0p} is nonvanishing only when $k$ and $k'$ 
belong to the same set, and $l$ and $l'$ also. It can be seen, 
by writing ${\bf \tilde G}_0^{(p)}$ in block diagonal form, that 
this property is also satisfied by the inverse matrix 
${\bf F}^{(p)} =  ({\bf \tilde G}_0^{(p)})^{-1}$. Another useful 
property of the matrices ${\bf \tilde G}_0^{(p)}$ is the following. 
We see that permuting $k$ and $l$, and $k'$ and $l'$ in \eqref{G0p}, 
is equivalent to complex conjugation. Consequently, 
$( {\bf \tilde G}_0^{(-p)} )_{lk,l'k'}=( {\bf \tilde G}_0^{(p)})^*_{kl,k'l'}$, 
and similarly for the inverses ${\bf F}^{(p)}$. At zero temperature, 
$( {\bf \tilde G}_0^{(p)} )_{kl,k'l'}$ vanishes if $E_{l}>E_{l'}$, 
and hence $( {\bf F}^{(p)} )_{kl,k'l'}=0$ for $E_{l}>E_{l'}$. Thus, 
in particular, $( {\bf F}^{(1)} )_{kl,c^- a}=\delta_{kc^-}\delta_{la}/
( {\bf \tilde G}_0^{(1)} )_{c^-a,c^-a}$.

\section{No thermal equilibrium Schmidt decomposable 
state}\label{app:NSdgs}

We consider a composite system consisting of ${\cal N}$ 
subsystems, with no long-range interaction between 
these subsystems. The Hamiltonian ${\cal H}$ of 
the complete system can thus be decomposed as 
${\cal H}={\cal H}_1+{\cal H}'$ where no observable of 
subsystem $1$ appears in ${\cal H}'$ and there exists 
a subsystem $n \ne 1$ which is not affected by ${\cal H}_1$. 
The Hamiltonian ${\cal H}$ is assumed to have 
a nondegenerate ground state $| 0 \rangle$. The thermal 
equilibrium state of the sytem is thus pure in the zero 
temperature limit. We show that $| 0 \rangle$ is not 
a Schmidt decomposable state 
$| \Psi \rangle=\sum_{r=1}^s \lambda_r 
\otimes_{n=1}^{\cal N} | \psi_n^{(r)} \rangle$ 
where $\langle \psi_n^{(r)} | \psi_n^{(r')} \rangle=\delta_{rr'}$, 
$s \in \{ 2, \ldots, \mathrm{min}_n(d_n) \}$, and $d_n$ is 
the dimension of the subsystem $n$ Hilbert space \cite{P,T}, 
as follows. Since $\langle \Phi_r | {\cal H}'  | \Phi_{r'} \rangle
\propto \langle \psi_1^{(r)} | \psi_1^{(r')} \rangle=\delta_{rr'}$ 
where $| \Phi_r \rangle = \otimes_{n=1}^N | \psi_n^{(r)} \rangle$, 
and similarly for ${\cal H}_1$, 
$A=\langle \Psi | {\cal H} | \Psi \rangle=\sum_r |\lambda_r|^2 
\langle \Phi_r | {\cal H}  | \Phi_{r} \rangle$. 
Consequently, $A > {\langle 0 | {\cal H} | 0 \rangle}$, 
as this last value is the minimum possible one for 
the average energy, and $|  0 \rangle$ is unique. 
Thus, $|  0 \rangle$ is not $|  \Psi \rangle$.

\section{No decoherence-free subspace}\label{app:Ndfs}

We show here that, even in the limiting case 
$\epsilon_f=\epsilon=0$, there is no decoherence-free 
subspace in the TLS Hilbert space. This can be directly 
proved from the corresponding Hamiltonian expression. 
For $\epsilon_f=\epsilon=0$, the Hamiltonian \eqref{H} 
simplifies to $H=H_{TLS}+H_{\cal B}+e\pi S_x$ 
where $S_x=\sum_n \sigma_n^x$, and hence, 
the coupling of the TLS to their environment, is 
described by a product term $\pi S_x$ where $\pi$ is 
an observable of the environment and $S_x$ is 
an observable of the TLS system. Thus, 
a decoherence-free subspace ${\cal S}$ would be 
a space spanned by eigenvectors of $S_x$ with 
the same eigenvalue and invariant under the TLS 
Hamiltonian $H_{TLS}$. In such a vector space 
${\cal S}$, there would exist a basis 
consisting of eigenstates of $H_{TLS}$. In other words, 
$H_{TLS}$ and $S_x$ would have common eigenvectors. 
Such a state is also eigenvector of $[H_{TLS},S_x]$ 
with eigenvalue $0$. 

For any state $| \psi \rangle = \sum_r \lambda_r 
\mathop {\hbox{\c\char'012}}_{n}| \eta^{(r)}_n \rangle_n$, 
the state $| \psi' \rangle=[ H_{TLS} , S_x ] | \psi \rangle$ 
is given by
\begin{equation}
| \psi' \rangle = 2 \sum_r \lambda_r 
\sum_{n=1}^{2N} \left[ J\eta_n^{(r)} 
\big(\eta_{n-1}^{(r)}+\eta_{n+1}^{(r)} \big) 
+ h_n \eta_n^{(r)} \right] \sigma_n^x |  
\mbox{\large $\eta$}_r \rangle 
\label{com}
\end{equation}
where $|  \mbox{\large $\eta$}_r \rangle 
= \mathop {\hbox{\c\char'012}}_{n}| \eta_n^{(r)} \rangle_n$, 
with the conventions $h_{2N+1-n}=-h_n$ and 
$\eta^{(r)}_{-1}=\eta^{(r)}_{2N+1}=0$. We now assume that 
$| \psi \rangle$ is an eigenstate of $H_{TLS}$. In this case, all 
the configuration states $|  \mbox{\large $\eta$}_r \rangle$ 
have the same energy. Since, for $h_n \ne 0$, no coefficient 
in the decomposition \eqref{com} vanishes, a term in 
this sum does not contribute only if it is the opposite 
of another one. This is only possible if 
there exist $n'$ and $r'$ such that 
$\sigma_{n'}^x |  \mbox{\large $\eta$}_{r'} \rangle
=\sigma_n^x |  \mbox{\large $\eta$}_r \rangle$. 
For general values of the fields $h_n$, 
$ |  \mbox{\large $\eta$}_{r'} \rangle
= \sigma_{n'}^x \sigma_n^x |  \mbox{\large $\eta$}_r \rangle$ 
and $|  \mbox{\large $\eta$}_r \rangle$ can have the same 
energy only if $n'=2N+1-n$, 
$\eta^{(r)}_{2N+1-n}=\eta^{(r)}_{n}$, and 
$\eta^{(r)}_{2N-n}+\eta^{(r)}_{2N-n+2}
=-\eta^{(r)}_{n-1}-\eta^{(r)}_{n+1}$ 
if $n \notin \{N,N+1\}$ replaced by 
$\eta^{(r)}_{N+2}=-\eta^{(r)}_{N-1}$ if 
$n=N$ or $N+1$. Consequently, for given $n$ and $r$, 
if $n'$ and $r'$ exist, they are unique. Consider now 
a given $r$. The conditions on $\mbox{\large $\eta$}_r$ 
found above, clearly show that there cannot exist 
$n_{-1}$, $r_{-1}$, $n_0$, $r_0$, $n_1$ 
and $r_1$ such that $\sigma_{n_\tau}^x |  
\mbox{\large $\eta$}_{r_\tau} \rangle
=\sigma_{n+\tau}^x |  \mbox{\large $\eta$}_r \rangle$. 
Thus, there always remains at least one non-vanishing 
$(\langle \mbox{\large $\eta$}_r | \sigma_n^x) 
[H_{TLS},S_x] | \psi \rangle$. 
Therefore, $[H_{TLS},S_x] | \psi \rangle \ne 0$, 
and hence $H_{TLS}$ and $S_x$ have no common 
eigenvector. There is thus no decoherence-free subspace 
in the TLS Hilbert space. This can happen in particular cases. 
For example, for $N=2$ and $h_2=0$, $({| +-++ \rangle} - 
{| --+- \rangle} - {| ++-+ \rangle} + {| -+-- \rangle})/2$ 
is eigenstate of both $H_{TLS}$ and $S_x$, and hence of 
$H=H_{TLS}+H_{\cal B}+e\pi S_x$. If the TLS are initially 
prepared in this state, they remain in it for ever.  

\section{Two-heat-reservoir environment}\label{app:Thre} 

In this Appendix, we show that the TLS asymptotic state 
$\rho_{\infty}$ can be a pure Schr\"odinger cat state if 
the monochromatic field is replaced by a second heat 
bath ${\cal B}'$. We consider the Hamiltonian 
\begin{equation}
H = H_{TLS}  + H_{\cal B}+ H_{{\cal B}'} 
+  e \sum_{n=1}^{2N}  \sigma^x_n \pi + 
\epsilon' \sum_{n=1}^{2N}  \sigma^x_n \pi_n
\end{equation}
where $\pi$ is an observable of bath ${\cal B}$ and 
$\pi_n$ are observables of bath ${\cal B}'$, 
and the initial state
\begin{equation}
\Omega = \sum_{k,l} r_{kl} | k \rangle \langle l |
\otimes Z^{-1} e^{ -H_\mathrm{\cal B}/T}
\otimes (Z')^{-1} e^{ -H_\mathrm{{\cal B}'}/T'}
\end{equation}
where $Z'=\mathrm{Tr}\exp(-H_\mathrm{{\cal B}'}/T)$ 
and $T'$ is the temperature of bath ${\cal B}'$. We are 
concernerd with the regime ${\epsilon \ll \epsilon' \ll e}$, 
which is the analog of that considered in section 
\ref{sec:SScs}. In this regime, the deviation of 
the coupling to bath ${\cal B}$ from the ideal form 
$S_x \pi$, can be neglected, and $\epsilon$ has been 
set to zero in the above expression of $H$. Here, 
$\rho_{\infty}$ is a steady state. From section 
\ref{sec:Uchb}, we know that it is of the form \eqref{rhos2}. 
As in section \ref{sec:Ltl}, it can be shown that the potential 
probabilities $p_q$ where $q \ne \pm$, vanish in the low 
temperature limit, faster than $\exp(-3J/T)$. The difference 
is simply that here the downward transitions are induced 
by bath ${\cal B}'$.

The ratio $p_+/p_-$ can be determined by a perturbative 
calculation similar to that done in section \ref{sec:SScs}. 
The state $\rho_\infty$ obeys 
$\sum_{k',l'} \gamma_{kl,k'l'} (i0^+) {u}^{(0)}_{k'l'} = 0$, 
where $k$ and $l$ are such that $E_k=E_l$, $k'$ and $l'$ 
are such that $E_{k'}=E_{l'}$, and $\gamma_{kl,k'l'} (i0^+)$ is 
given by an expression of the form \eqref{gammareel} but with 
sums over eigenstates of both baths ${\cal B}$ and ${\cal B}'$. 
This equation set can be written in matrix form as 
$[{e^2 {\bf G}}+{(\epsilon')^2 {\bf G}'}]{\bf u}=0$ where 
${\bf G}$ is the matrix ${\bf \tilde G}_0^{(0)}$ introduced in 
section \ref{sec:SScs}. The matrix term proportional to 
$e\epsilon'$, vanishes under the assumptions 
$\mathrm{Tr} (\Omega \pi)=\mathrm{Tr} (\Omega \pi_n^x)=0$. 
Using the left and right eigenvectors of ${\bf G}$, 
${\bf \Phi}_q$ and ${\bf \Psi}_q$, it can be shown that 
$\sum_{q'} r_{q q'} p_{q'} = 0$. Here the rates $r_{qq'}$ are 
given by, for $q \ne q'$,
\begin{multline}
r_{qq'}= \frac{2\pi}{Z' z_{q'}} 
\sum_{{k \in{\cal E}_q \atop l\in{\cal E}_{q'}} } 
\sum_{A,B} e^{-E_A/{T'}} e^{-E_l/T} 
\delta(E_A-E_B+\omega_{lk})  \\ 
\times \Big| \sum_{n,\nu}
\langle l | \sigma^x_n | k \rangle 
\langle A | \pi_n^x  | B \rangle \Big|^2 , \label{r2}
\end{multline}
where $A$ and $B$ run over the eigenstates of 
bath ${\cal B}'$, and by $r_{q'q'}=-\sum_{q\ne q'} r_{qq'}$. 
All transitions, induced by bath ${\cal B}'$, 
from a state $| l \rangle$ given by ${\cal E}_{q'}$ to a state 
$| k \rangle$ given by ${\cal E}_{q}$, contribute to \eqref{r2}. 
For $T'=T$, $z_{q'}r_{qq'}=z_q r_{q'q}$, which ensures that 
the equilibrium density matrix 
$\rho_{eq} \propto \exp(-H_{TLS}/T)$ 
is a steady state of the TLS system. 
We consider the low temperature regime 
$T' \ll T \ll  h_n < J$. 
In the sum \eqref{r2}, the terms such that $E_l < E_k$, 
vanish in the zero $T'$ limit, whereas the terms such that 
$E_l > E_k$, reach finite values. Thus, in the regime 
considered, the upward transitions induced by bath ${\cal B}'$ 
are negligible and only the downward transitions contribute 
to \eqref{r2}. Consequently, $r_{+-} \sim \exp[-(E_c-E_0)/T]$, 
$r_{++} \simeq -r_{-+} \ll  r_{+-}$ (for $N \ge 3$), and hence, 
since $p_q \ll \exp(-3J/T)$ for $q \ne \pm$, $p_-$ vanishes, 
and the TLS steady state is $| \mathrm{Scs}^+ \rangle$.

\end{appendix}

\end{document}